\documentclass{emulateapj}
\usepackage{amsmath,color}

\newcommand{\rate}{${\rm Gpc}^{-3}{\rm yr}^{-1}$}

\begin{document}

\title{Implications of binary black hole detections on the merger rates
of double neutron stars and neutron star-black holes}

\author{Anuradha Gupta}
\email{axg645@psu.edu}
\affiliation{Institute for Gravitation and Cosmos, Physics Department, Pennsylvania State University, University Park, PA, 16802, USA}

\author{K. G. Arun}
\email{kgarun@cmi.ac.in}
\affiliation{Chennai Mathematical Institute, Siruseri, India}
\affiliation{Institute for Gravitation and Cosmos, Physics Department, Pennsylvania State University, University Park, PA, 16802, USA}

\author{B. S. Sathyaprakash}
\email{bss25@psu.edu}
\affiliation{Institute for Gravitation and Cosmos, Physics Department, Pennsylvania State University, University Park, PA, 16802, USA}
\affiliation{Department of Astronomy \& Astrophysics, Pennsylvania State University, University Park, PA, 16802, USA}
\affiliation{School of Physics and Astronomy, Cardiff University, 5, The Parade, Cardiff, UK, CF24 3AA}

\begin{abstract}
We show that the inferred merger rate and chirp masses of binary black holes (BBHs) detected by advanced LIGO (aLIGO) can be used to constrain the rate of double neutron star (DNS) and neutron star - black hole (NSBH) mergers in the universe. We explicitly demonstrate this by considering a set of publicly available population synthesis models of \citet{Dominik:2012kk} and show that if all the BBH mergers, GW150914, LVT151012, GW151226, and GW170104, observed by aLIGO arise from isolated binary evolution, the predicted DNS merger rate may be constrained to be $2.3-471.0$~\rate~ and that of NSBH mergers will be constrained to  $0.2-48.5$~\rate. The DNS merger rates are not constrained much but the NSBH rates are tightened by a factor of $\sim 4$ as compared to their previous rates. Note that these constrained DNS and NSBH rates are extremely model dependent and are compared to the unconstrained values $2.3-472.5$ \rate~ and $0.2-218$ \rate, respectively, using the same models of \citet{Dominik:2012kk}.
These rate estimates may have implications for short Gamma Ray Burst progenitor models assuming they are powered (solely) by DNS or NSBH mergers.  
While these results are based on a set of open access population
synthesis models which may not necessarily be the representative ones,
the proposed method is very general and can be applied to any number of
models thereby yielding more realistic constraints on the DNS and NSBH merger rates from the inferred BBH merger rate and chirp mass. 
\end{abstract}


\section{Introduction}
The first three binary black hole (BBH) detections by the 
aLIGO detectors~\citep{GW150914,GW151226,GW170104} have fundamentally impacted our
understanding of the astrophysics and underlying physics \citep{Rates,
Astro, TOG, O1BBH}. With these detections, the rate of 
BBH mergers is constrained between $12-213$ \rate~ \citep{GW170104}. It is
expected that, as aLIGO and Virgo reach their design
sensitivity, we will detect double neutron star (DNS) and neutron
star--black hole (NSBH) mergers too. In fact, LIGO Scientific Collaboration had predicted DNS and NSBH merger rates to be $10-10000$ \rate~ and $0.5-1000$ \rate~ respectively by extrapolating the rates from the observed binary pulsars in the Milky Way Galaxy, way before their first observing run had begun (see \citet{LSCrates} and references therein). After the non-detection of DNS and NSBH coalescences in the first observing run, they placed upper bounds on the DNS and NSBH merger rates to be $12600$ \rate~ and $3600$ \rate, respectively \citep{LIGOBNSrates}.

Many population synthesis and Monte Carlo simulations
have also predicted the rates of DNS and NSBH systems in our universe \citep{Tutukov1993, Hurley:2002rf, Voss:2003ep, Nelemans:2003xp, Belczynski_2002, Belczynski_2008, Dominik:2012kk, Dominik:2013tma}. However, these
rates are highly uncertain and span three orders of
magnitude,  depending on the assumptions that go into these
simulations \citep{Dominik:2012kk}. While certain population synthesis models do not have the
capability to predict the rates of all three compact binary populations (DNS, NSBH and BBH), 
some of them can do so  for a given set of input
parameters \citep{Belczynski_2002, Belczynski_2008, Gultekin2004, 
OLeary2006, Grindlay2006,Sadowski2008,Ivanova:2007bu,Downing2010, Miller:2008yw}. Typically, such simulations consider many different models
which capture different input parameters to model the galaxy and
physics associated with the binary evolution and obtain the rates of all
three binary populations. The input
parameters may include metallicity of the stellar environment, mass loss
due to stellar winds, details of mass transfer episodes in the binary
evolution, kick imparted by the supernova explosion, chemical homogeneity of the surrounding, and age of galactic globular clusters if binaries are formed in dense clusters via dynamical interactions. It is interesting to note that different input physics and formation channels lead to very different merger rates of compact binaries containing NSs. For instance, models involving chemically homogeneous evolution \citep{Mandel:2015qlu} or dynamical interactions \citep{Gultekin2004, 
OLeary2006, Grindlay2006,Sadowski2008,Ivanova:2007bu,Downing2010, Miller:2008yw} are tend to produce very less DNS and NSBH mergers as compared to the scenarios where binaries form in isolation \citep{Dominik2012}.  

In this work, we argue that we should be able to combine the results of those population synthesis models which can
predict the rates of all three compact binary mergers (DNS, NSBH and BBH) with the
{\it inferred} BBH merger rates by aLIGO and chirp masses\footnote{Chirp mass is the best measured parameter by gravitational wave observations during the inspiral phase of a binary and defined as ${\cal M} = (m_1m_2)^{3/5}/(m_1+m_2)^{1/5}$, where $m_1$ and $m_2$ are the component masses.} of the detected BBHs, 
leading to a considerable reduction in the uncertainty in the
predicted DNS and NSBH merger rates. Intuitively, the inferred BBH
merger rates and their properties can constrain the uncertain physics which go
into the population synthesis models thereby narrowing down the range
of DNS and NSBH merger rates. Indeed, this method implicitly assumes
that the set of models we use to carry out this study is representative
of the actual binaries that exist in nature. Our rate estimates can hence be
refined when more accurate and more representative models are available.
We note that several other methods have been proposed in the past to compare 
and constrain various astrophysical binary formation models using the compact 
binary merger rates and their observed parameters~\citep{Mandel:2015spa, 
Messenger:2012jy, OShaughnessy:2012oew, Mandel:2009nx, Mandel:2010xq, 
Dominik2015, Stevenson:2015bqa, Zevin2017}. Similar studies has also been done 
using supernovae rates \citep{Richard_2008} and DNS populations 
\citep{Richard2005ApJ}.

The rest of the paper is organized as follow: In Sec.~\ref{models} we briefly 
describe the binary formation models of \citet{Dominik:2012kk} which predict 
merger rates for all the three binary populations. Section~\ref{method} explains 
our method of constraining DNS/NSBH merger rates while making use of BBH 
detections from aLIGO.  Section~\ref{discussion} discusses the implications 
of our findings.

\section{Isolated binary formation models}
\label{models}
\citet{Dominik:2012kk} (hereafter Dominik2012) proposed formation models for `isolated' compact binaries (DNS, NSBH and BBH) merging in a Hubble time
and calculated their merger rates using the StarTrack population synthesis code \citep{Belczynski_2002, Belczynski_2008}.
Moreover, these models are publically available and provide distribution of chirp masses of the compact binary and other properties of the mergers \footnote{http://www.syntheticuniverse.org}.  Since the details of the common envelope, the maximum mass of the NS, physics of supernova explosions forming compact objects and wind mass loss rates of the progenitor stars are still very uncertain, Dominik2012 provided a set of population synthesis models while changing the associated parameters and input physics one at a time. There are 16 models: 1 standard (S) and 15 variations (V1-V15), and all of them are summarized in Table I in Dominik2012. In the standard model, the maximum NS mass is assumed to be $2.5M_{\odot}$, the rapid supernova engine \citep{Fryer2012} and a physically motivated common envelope binding energy \citep{Xu_Li_2010} have been used while the natal kicks in core-collapse supernovae are drawn from a Maxwellian distribution with $\sigma = 265$ km/s \citep{Hobbs_2005}.

Additionally, Dominik2012 considered two scenarios for the Hertzsprung gap of the donar stars. In the first one they ignore the core-envelope boundary issue and calculated common envelope energetics as normal.  This is the {\it optimistic} Hertzsprung gap evolution model (submodel A). In the other scenario they assume that when a binary, in which the donor star is on the Hertzsprung gap, enters into a common envelope phase, merges prematurely. This reduces the number of merging compact binaries and hence their rates. This is the {\it pessimistic} Hertzsprung gap evolution model (submodel B). Dominik2012 also considered two fiducial stellar populations with different metallicity: solar ($Z=Z_{\odot}$) and sub-solar ($Z=0.1Z_{\odot}$). Consequently, they have models for four scenarios: (1) submodel A, solar, (2) submodel B, solar, (3) submodel A, sub-solar, and (4) submodel B, sub-solar. Hence, in total, there are 64 formation models. The rates and chirp mass estimates for DNS, NSBH and BBH mergers predicted by all these 64 models are given in Tables 2-3 and 6-9 in Dominik2012. 
If we assume these 64 models to be the representative ones in estimating the merger rates of compact binaries, we find that DNSs will have merger rates in $0.6-774$ \rate~ while the merger rates for NSBHs will be $0-330$ \rate. 

However, in reality, the universe has a distribution of metallicities which is also a function of redshift: low-metallicity at high redshift and vice-versa. The metallicity of the stellar environment affects the merger rates of compact binaries especially those containing BHs with lower metallicity ——leading to higher binary merger rates \citep{Belczynski2010ApJ}. Therefore, the above DNS and NSBH merger rates predicted by Dominik2012 using all 64 models may not be realistic as they are derived from only two metallicities. To overcome this issue, we instead use rate estimates for DNS, NSBH and BBH systems as a $50\%-50\%$ contribution from both low and high metallicity environments \citep{Belczynski2010ApJ, Stevenson:2015bqa}. We take the average of solar and sub-solar metallicity rates as a rough approximation, predicted by an underlying model, {\it i.e.,} submodel A and submodel B. In this way, we now have only 32 models (16 submodel A and 16 submodel B models) with DNS and NSBH merger rates to be $2.3-472.5$ \rate~ and $0.2-218$ \rate, respectively,  as compared to $0.6-774$ \rate~ and $0-330$ \rate. Similarly, one can combine the chirp mass estimates from solar and sub-solar models by taking the minimum (maximum) of minimum (maximum) chirp masses predicted by  solar and sub-solar models. We will use these combined rates and chirp masses from solar and sub-solar scenarios to rule out the binary formation models.

We note that later on \citet{Dominik2015} incorporated the effect of cosmological evolution of binaries in their StarTrack code, and computed merger rates for DNS, NSBH and BBH systems. They considered two scenarios for metallicity evolution: (1) high end: a distribution of metallicities with median $1.5Z_{\odot}$ at redshift $z=0$, (2) low end: $50\%$ of the stars form in galaxies at $z\sim0$ with $Z=0.2Z_{\odot}$ whereas the other $50\%$ have metallicity $Z=1.5Z_{\odot}$.
Further, in \citet{deMink:2015yea} it has been noted that the treatment for tidal locking of stars in Dominik2012 was incorrect which led to an over estimation of compact binary merger rates. We are, however, not considering these models in our analysis.

Soon after the detection of GW150914, \citet{Belczynski:2016obo} updated their StarTrack code and proposed a new set of models to interpret the origin of GW150914 from the evolution of isolated binaries. Although the proposed models of Belczynski {\it et al.} are more realistic, the authors have not made the predicted merger rates for DNS and NSBH systems public. Similarly,  while using the rapid binary population synthesis code COMPAS, \citet{Stevenson:2017tfq} demonstrated that the origin of the first three BBHs can be explained by classical isolated binary evolution in a low-metallicity environment. These models as well do not provide any estimate for DNS and NSBH merger rates. Therefore, in our analysis, we do not consider these models to constrain the DNS and NSBH merger rates.  

Very recently, \citet{Belczynski:2017gds} proposed another suite of population synthesis models to explain the origin of GW170104-like BBHs as there could be a misalignment between the orbital angular momentum of the binary and BH spins \citep{GW170104} assuming that the spin magnitudes are {\it not} inherently small \citep{Farr:2017uvj}.   
These models suggest that all the BBHs detected by aLIGO so far can be formed from isolated binary evolution with BHs having small natal spins.
Unlike \citet{Belczynski:2016obo}, \citet{Belczynski:2017gds} provide the merger rates of DNS and NSBH systems along with that of BBHs.
Our constraints on DNS and NSBH  merger rates will be unaffected even if we include these latest models in our analysis.

\citet{Chruslinska2017} recently demonstrated that the StarTrack code used in Dominik2012 does not incorporate enough physics regarding the natal kick velocities of NSs in DNSs and hence fails to reproduce orbital period and eccentricities of the observed galactic DNSs. While improving upon the natal kick distribution of NSs in the DNS population, \citet{Chruslinska2017} also provide galactic merger rates for DNSs. We verify that the DNSs merger rates predicted by \citet{Chruslinska2017} fall within the bounds provided in this paper.  

In the next section, we show how we can tighten the DNS and NSBH merger rate estimates with the help of known BBH mergers.

\section{Methodology}
\label{method}
After the third aLIGO detection, GW170104, the BBH merger rates are revised as $12-213$ \rate~ \citep{GW170104}. Moreover, 
after the end of the first observing run, aLIGO has also placed an upper bound on the DNS and NSBH merger rates as  $12600$ Gpc$^{-3}$yr$^{-1}$ and $3600$ Gpc$^{-3}$yr$^{-1}$, respectively \citep{LIGOBNSrates}. In this paper, we present
an independent method of constraining the DNS and NSBH merger rates from the
BBH detections by aLIGO.

We rule out those isolated binary formation models which predict BBH merger rates outside the range inferred by aLIGO ($12-213$ \rate). In this way, we rule out $14$ models and the remaining $18$ models are given in Table~\ref{Tab_rate_cut}. For DNS binaries, the lowest merger rate is predicted by both submodel A and submodel B scenarios of V1 model ($2.3$ \rate)~ whereas the highest DNS merger rate is given by submodel A scenario of V15 model ($471$ \rate). Note that we use the following conversion to translate the rates given in Dominik2012 in the units of \rate,

\begin{equation}
1 \, {\rm MWEG}^{-1} {\rm Myr}^{-1} = 10 \, {\rm Gpc}^{-3} {\rm yr}^{-1}\,,
\end{equation}
where MWEG stands for Milky Way Equivalent Galaxy. The resulting bound on DNS merger rates, $2.3-471$ \rate, is not much different than the previous rates. 
Similarly, the surviving models place a bound on predicted NSBH merger rate to be $0.2-170$ \rate, which has tightened by a factor of $\sim 1.3$.

\begin{table}[h]
\caption{The isolated binary formation models that survived after the BBH rates constraint.}
\label{Tab_rate_cut}
\begin{center}
\begin{tabular}{c|c} 
scenario & models\\
\hline
submodel A & V1, V4, V8, V15 \\
submodel B & S, V1, V2, V3, V5, V6, V7, V9  \\
& V10, V11, V12, V13, V14, V15 \\
\hline
\end{tabular}
\end{center}
\end{table}

\begin{table}[h]
\caption{$90\%$ credible bounds on the chirp mass for 3 GW detections, GW150914 \citep{GW150914}, GW151226 \citep{GW151226}, GW170104 \citep{GW170104} and 1 candidate LVT151012 \citep{O1BBH}.}
\label{Tab_mc}
\begin{center}
\begin{tabular}{c|c} 
GW event & Chirp mass range ($M_{\odot}$)\\
\hline
GW150914 & 26.6-29.9 \\ 
LVT151012 & 14.0-16.5 \\
GW151226 & 8.6-9.2 \\
GW170104 & 18.4-23.5 \\
\hline
\end{tabular}
\end{center}
\end{table}

We can also rule out certain binary formation models on the basis of chirp mass measurements of detected BBHs. In Table~\ref{Tab_mc}, we provide the $90\%$ credible bound on each of the GW events and we see that the lowest and the highest observed chirp masses are $8.6M_{\odot}$ and  $29.9M_{\odot}$, respectively. As there are uncertainties in the chirp mass measurement, we take a rather conservative approach in considering the lowest and highest measured chirp masses to rule out the isolated binary formation models. We use the $90\%$ upper limit for GW151226 ($9.2M_{\odot}$) as our lowest measured chirp mass and the $90\%$ lower limit for GW150914 ($26.6M_{\odot}$) as the highest measured chirp mass.
Therefore, we rule out models whose lowest estimated chirp mass $>9.2M_{\odot}$ or highest estimated chirp mass $<26.6M_{\odot}$. In this way, we rule out $14$ models and the remaining ones are listed in Table~\ref{Tab_mc_cut}. We find that the surviving models in Table~\ref{Tab_mc_cut} predict the DNS and NSBH merger rates to be $2.3-472.5$ \rate~ and $0.2-218$ \rate, respectively. This implies that ruling out the binary formation models on the basis of observed BBH chirp masses does not constrain DNS and NSBH merger rates. Note that the predicted merger rate for NSBHs would have been tighter ($0.2-94.5$ \rate), if we have had chosen $8.6M_{\odot}$ and $29.9M_{\odot}$ as our lowest and highest observed chirp masses.

\begin{table}[h]
\caption{The isolated binary formation models that survived after the BBH chirp mass constraint.}
\label{Tab_mc_cut}
\begin{center}
\begin{tabular}{c|c} 
scenario & models\\
\hline
submodel A & S, V1, V2, V3, V4, V5, V6, V7, V8, \\
& V9, V10, V11, V12, V13, V14, V15 \\
submodel B & V1, V15 \\
\hline
\end{tabular}
\end{center}
\end{table}

Finally, if we apply both the constraints, {\it i.e.,} BBH rates and chirp masses, only $6$ out of $32$ models survived and they are listed in Table~\ref{Tab_both_cuts}. 
We find that the lowest merger rate for DNS systems ($2.3$ Gpc$^{-3}$ yr$^{-1}$) is given by both the submodel A and submodel B scenarios of V1 model whereas the highest rate ($471$ Gpc$^{-3}$ yr$^{-1}$) is predicted by submodel A scenario of V15 model. Therefore, the final bound on the DNS merger rate after both the constraints is $2.3-471$ Gpc$^{-3}$yr$^{-1}$. Similarly, the final bound on the NSBH merger rates turned out to be $0.2-48.5$ Gpc$^{-3}$yr$^{-1}$. Our method has tightened the DNS merger rates slightly whereas that of NSBHs are tightened by a factor of $\sim 4$. The predicted NSBH merger rates would have been $0.2-3.6$ \rate~, tightened by a factor of $\sim 60$, if we had chosen $8.6M_{\odot}$ and $29.9M_{\odot}$ as our lowest and highest observed chirp masses.

\begin{table}[h]
\caption{The isolated binary formation models that survived after both BBH rates and chirp mass constraints.}
\label{Tab_both_cuts}
\begin{center}
\begin{tabular}{c|c} 
scenario & models\\
\hline
submodel A & V1, V4, V8, V15 \\
submodel B & V1, V15 \\
\hline
\end{tabular}
\end{center}
\end{table}

\section{Discussions}
\label{discussion}
In this paper, we have shown that it is possible to constrain DNS and NSBH merger rates from the BBH detections by aLIGO. Many of the population synthesis models predict merger rates and other properties (such as chirp mass) for all three types of binary populations (DNS, NSBH and BBH), assuming similar physics and input parameters. We argue that those models which predict BBH merger rates and chirp masses outside the range inferred by aLIGO will be ruled out. Consequently, the predicted DNS/NSBH merger rates inferred from the remaining models are better constrained. As a demonstration, we applied this method on publicly available models of Dominik2012. Assuming all the BBHs, GW150914, LVT151012, GW151226, and GW170104, observed by aLIGO are formed from isolated binary evolution, we find that the DNS merger rate will be constrained to $2.3-471$ \rate~  whereas the NSBH mergers will have rates in $0.2-48.5$ \rate. Note that the DNS and NSBH merger rates predicted by the full set of models by Dominik2012 were $2.3 - 472.5$ \rate~ and $0.2-218$ \rate, respectively, before such constraint was applied. Therefore, our method has tightened the NSBH merger rates by a factor of $\sim 4$ whereas the DNS merger rates are marginally constrained. 
While the results presented here are limited to a specific formation channel, the method proposed in this paper can be applied to any number of population synthesis models. 

Our underlying  assumption of all the detected BBHs being formed via isolated binary evolution provides the highest possible upper bound on the DNS and NSBH merger rates. This is because if the BBHs would have formed from some other channels, e.g., dynamical formation, then the predicted merger rates for DNSs/NSBHs will be even smaller for those models.

Note that a fundamental assumption of this {\it Letter} is that BBHs and DNSs/NSBHs form from the same type of progenitors, namely isolated massive binary systems and the same underlying physical mechanism. In reality, these two populations can originate from multiple progenitors some of which could be common to both and others distinct to a specific system. Hence, it is possible that the merger rates of these systems are not quite related. Future studies would need to consider multiple formation scenarios while constraining rates of different populations.

It is interesting to note that the new rates we quote here based on the
BBH detections are much stronger than the upper
limits from the actual DNS/NSBH searches ~\citep{LIGOBNSrates} by
aLIGO, though the methods are very different. 
Our new rate estimates may have implications for SGRB progenitor
models assuming they are powered (solely) by DNS or NSBH mergers. For
instance, after accounting for the beaming effect, the rate of  observed
SGRBs, $3-30$ \rate~ \citep{Coward:2012gn}, can not be accounted for only by a population of NSBH mergers whereas the DNS merger
rates may account for SGRBs with an appropriate beaming correction. The beaming angle for SGRBs is given as

\begin{equation}
\cos \theta_j = 1-\frac{R_{\rm SGRB}}{R_{\rm merger}} \,,
\end{equation}
where $R_{\rm SGRB}$ is the SGRB rate and $R_{\rm merger}$ is the progenitor merger rate.  Therefore, with the new predicted merger rates for DNSs (NSBHs) one can place a bound on the beaming angle to be $6.5^{\circ}-107.7^{\circ}$ ($20.3^{\circ}-67.6^{\circ}$) assuming SGRBs are solely produced by DNS (NSBH) mergers. On the other hand, if we consider NSBH merger rates from pessimistic chirp mass constraints, {\it i.e.}, $0.2-3.6$ \rate, one can place only a lower bound on the beaming angle of $80.4^{\circ}$, assuming SGRBs are solely produced by NSBH mergers. This lower bound on beaming angle seems unrealistic for NSBH mergers to be the only progenitors of SGRBs.

Any future BBH detections will further tighten the DNS/NSBH merger rates as it will not only place a tighter bound on the current BBH merger rates but will also improve observed chirp mass distribution. For example, let us consider that, in future, aLIGO detects a BBH merger of chirp mass $< 5M_{\odot}$ and BBH merger rates are updated to be $25-150$ \rate. Then both submodel A and submodel B scenarios of V1 model will be ruled out and the resulting DNS and NSBH merger rates will be $128-471$ \rate~ and $1.5-48.5$ \rate, respectively, which would tighten at the lower end side.

As mentioned earlier, the models of Dominik2012 used in this paper are not the representative ones and have many shortcomings as compared to the models proposed recently \citep{Belczynski:2016obo, Stevenson:2017tfq, Belczynski:2017gds, Chruslinska2017}. Though these recent models are more realistic, there  are still many uncertainties involved which need to addressed. For example, these uncertainties include initial conditions \citep{deMink:2015yea}, modeling of massive stellar evolution, chemically homogeneous evolution, rotation of stars, magnetic fields and their effect on stellar wind strength \citep{Petit2017}, mass transfer efficiency and mass loss modes, common envelope binding energy, metallicity specific star formation rate as a function of redshift, and cosmological effects. The improvements in these binary formation models not only bring down the ariori uncertainties in the predicted DNS/NSBH  merger rate but can be further constrained using the method proposed in this paper.

\section*{Acknowledgments}
We thank Chad Hanna 
for useful discussions. 
KGA acknowledges the grant EMR/2016/005594 by Science and Engineering
Research Board (SERB), India. KGA is partially supported by a grant from Infosys foundation.
The authors acknowledge the support by the Indo-US Science and Technology Forum through the Indo-US Centre for the Exploration of Extreme
Gravity (IUSSTF/JC-029/2016).  This paper has the LIGO document number P1700209.


\begin{thebibliography}{}
\expandafter\ifx\csname natexlab\endcsname\relax\def\natexlab#1{#1}\fi

\bibitem[{Abadie {et~al.}(2010)}]{LSCrates}
Abadie, J., {et~al.} 2010, Class.Quant.Grav., 27, 173001

\bibitem[{Abbott {et~al.}(2016{\natexlab{a}})}]{Astro}
Abbott, B.~P., {et~al.} 2016{\natexlab{a}}, Astrophys. J., 818, L22

\bibitem[{Abbott {et~al.}(2016{\natexlab{b}})}]{O1BBH}
---. 2016{\natexlab{b}}, Phys. Rev., X6, 041015

\bibitem[{Abbott {et~al.}(2016{\natexlab{c}})}]{GW151226}
---. 2016{\natexlab{c}}, Phys. Rev. Lett., 116, 241103

\bibitem[{Abbott {et~al.}(2016{\natexlab{d}})}]{GW150914}
---. 2016{\natexlab{d}}, Phys. Rev. Lett., 116, 061102

\bibitem[{Abbott {et~al.}(2016{\natexlab{e}})}]{TOG}
---. 2016{\natexlab{e}}, Phys. Rev. Lett., 116, 221101

\bibitem[{Abbott {et~al.}(2016{\natexlab{f}})}]{Rates}
---. 2016{\natexlab{f}}, arXiv:1602.03842

\bibitem[{Abbott {et~al.}(2016{\natexlab{g}})}]{LIGOBNSrates}
---. 2016{\natexlab{g}}, Astrophys. J., 832, L21

\bibitem[{Abbott {et~al.}(2017)}]{GW170104}
---. 2017, Phys. Rev. Lett., 118, 221101

\bibitem[{{Belczynski} {et~al.}(2010){Belczynski}, {Dominik}, {Bulik},
  {O'Shaughnessy}, {Fryer}, \& {Holz}}]{Belczynski2010ApJ}
{Belczynski}, K., {Dominik}, M., {Bulik}, T., {et~al.} 2010, \apjl, 715, L138

\bibitem[{Belczynski {et~al.}(2016)Belczynski, Holz, Bulik, \&
  O'Shaughnessy}]{Belczynski:2016obo}
Belczynski, K., Holz, D.~E., Bulik, T., \& O'Shaughnessy, R. 2016, Nature, 534,
  512

\bibitem[{{Belczynski} {et~al.}(2002){Belczynski}, {Kalogera}, \&
  {Bulik}}]{Belczynski_2002}
{Belczynski}, K., {Kalogera}, V., \& {Bulik}, T. 2002, \apj, 572, 407

\bibitem[{{Belczynski} {et~al.}(2008){Belczynski}, {Kalogera}, {Rasio}, {Taam},
  {Zezas}, {Bulik}, {Maccarone}, \& {Ivanova}}]{Belczynski_2008}
{Belczynski}, K., {Kalogera}, V., {Rasio}, F.~A., {et~al.} 2008, apjs, 174, 223

\bibitem[{Belczynski {et~al.}(2017)}]{Belczynski:2017gds}
Belczynski, K., {et~al.} 2017, arXiv:1706.07053

\bibitem[{{Chruslinska} {et~al.}(2017){Chruslinska}, {Belczynski}, {Bulik}, \&
  {Gladysz}}]{Chruslinska2017}
{Chruslinska}, M., {Belczynski}, K., {Bulik}, T., \& {Gladysz}, W. 2017, ACTA
  ASTRONOMICA, 67, 37

\bibitem[{Coward {et~al.}(2012)Coward, Howell, Piran, Stratta, Branchesi,
  Bromberg, Gendre, Burman, \& Guetta}]{Coward:2012gn}
Coward, D., Howell, E., Piran, T., {et~al.} 2012, Mon. Not. Roy. Astron. Soc.,
  425, 1365

\bibitem[{de~Mink \& Belczynski(2015)}]{deMink:2015yea}
de~Mink, S.~E., \& Belczynski, K. 2015, Astrophys. J., 814, 58

\bibitem[{Dominik {et~al.}(2012{\natexlab{a}})Dominik, Belczynski, Fryer, Holz,
  Berti, Bulik, Mandel, \& O'Shaughnessy}]{Dominik:2012kk}
Dominik, M., Belczynski, K., Fryer, C., {et~al.} 2012{\natexlab{a}}, Astrophys.
  J., 759, 52

\bibitem[{Dominik {et~al.}(2012{\natexlab{b}})Dominik, Belczynski, Fryer, Holz,
  Berti, Bulik, Mandel, \& O'Shaughnessy}]{Dominik2012}
---. 2012{\natexlab{b}}, Astrophys. J., 759, 52

\bibitem[{Dominik {et~al.}(2013)Dominik, Belczynski, Fryer, Holz, Berti, Bulik,
  Mandel, \& O'Shaughnessy}]{Dominik:2013tma}
---. 2013, Astrophys. J., 779, 72

\bibitem[{{Dominik} {et~al.}(2015){Dominik}, {Berti}, {O'Shaughnessy},
  {Mandel}, {Belczynski}, {Fryer}, {Holz}, {Bulik}, \&
  {Pannarale}}]{Dominik2015}
{Dominik}, M., {Berti}, E., {O'Shaughnessy}, R., {et~al.} 2015, \apj, 806, 263

\bibitem[{{Downing} {et~al.}(2010){Downing}, {Benacquista}, {Giersz}, \&
  {Spurzem}}]{Downing2010}
{Downing}, J.~M.~B., {Benacquista}, M.~J., {Giersz}, M., \& {Spurzem}, R. 2010,
  Mon. Not. Roy. Astron. Soc., 407, 1946

\bibitem[{Farr {et~al.}(2017)Farr, Stevenson, Coleman~Miller, Mandel, Farr, \&
  Vecchio}]{Farr:2017uvj}
Farr, W.~M., Stevenson, S., Coleman~Miller, M., {et~al.} 2017, arXiv:1706.01385

\bibitem[{{Fryer} {et~al.}(2012){Fryer}, {Belczynski}, {Wiktorowicz},
  {Dominik}, {Kalogera}, \& {Holz}}]{Fryer2012}
{Fryer}, C.~L., {Belczynski}, K., {Wiktorowicz}, G., {et~al.} 2012, \apj, 749,
  91

\bibitem[{{Grindlay} {et~al.}(2006){Grindlay}, {Portegies Zwart}, \&
  {McMillan}}]{Grindlay2006}
{Grindlay}, J., {Portegies Zwart}, S., \& {McMillan}, S. 2006, Nature Physics,
  2, 116

\bibitem[{{G{\"u}ltekin} {et~al.}(2004){G{\"u}ltekin}, {Miller}, \&
  {Hamilton}}]{Gultekin2004}
{G{\"u}ltekin}, K., {Miller}, M.~C., \& {Hamilton}, D.~P. 2004, \apj, 616, 221

\bibitem[{{Hobbs} {et~al.}(2005){Hobbs}, {Lorimer}, {Lyne}, \&
  {Kramer}}]{Hobbs_2005}
{Hobbs}, G., {Lorimer}, D.~R., {Lyne}, A.~G., \& {Kramer}, M. 2005, Mon. Not.
  Roy. Soc., 360, 974

\bibitem[{Hurley {et~al.}(2002)Hurley, Tout, \& Pols}]{Hurley:2002rf}
Hurley, J.~R., Tout, C.~A., \& Pols, O.~R. 2002, Mon. Not. Roy. Astron. Soc.,
  329, 897

\bibitem[{Ivanova {et~al.}(2008)Ivanova, Heinke, Rasio, Belczynski, \&
  Fregeau}]{Ivanova:2007bu}
Ivanova, N., Heinke, C., Rasio, F.~A., Belczynski, K., \& Fregeau, J. 2008,
  Mon. Not. Roy. Astron. Soc., 386, 553

\bibitem[{Mandel \& de~Mink(2016)}]{Mandel:2015qlu}
Mandel, I., \& de~Mink, S.~E. 2016, Mon. Not. Roy. Astron. Soc., 458, 2634

\bibitem[{Mandel {et~al.}(2015)Mandel, Haster, Dominik, \&
  Belczynski}]{Mandel:2015spa}
Mandel, I., Haster, C.-J., Dominik, M., \& Belczynski, K. 2015, Mon. Not. Roy.
  Astron. Soc., 450, L85

\bibitem[{Mandel {et~al.}(2010)Mandel, Kalogera, \&
  O'Shaughnessy}]{Mandel:2010xq}
Mandel, I., Kalogera, V., \& O'Shaughnessy, R. 2010, 1637

\bibitem[{Mandel \& O'Shaughnessy(2010)}]{Mandel:2009nx}
Mandel, I., \& O'Shaughnessy, R. 2010, Class. Quant. Grav., 27, 114007

\bibitem[{Messenger \& Veitch(2013)}]{Messenger:2012jy}
Messenger, C., \& Veitch, J. 2013, New J. Phys., 15, 053027

\bibitem[{Miller \& Lauburg(2009)}]{Miller:2008yw}
Miller, M.~C., \& Lauburg, V.~M. 2009, Astrophys. J., 692, 917

\bibitem[{Nelemans(2003)}]{Nelemans:2003xp}
Nelemans, G. 2003, AIP Conf. Proc., 686, 263, [,263(2003)]

\bibitem[{{O'Leary} {et~al.}(2006){O'Leary}, {Rasio}, {Fregeau}, {Ivanova}, \&
  {O'Shaughnessy}}]{OLeary2006}
{O'Leary}, R.~M., {Rasio}, F.~A., {Fregeau}, J.~M., {Ivanova}, N., \&
  {O'Shaughnessy}, R. 2006, \apj, 637, 937

\bibitem[{O'Shaughnessy(2013)}]{OShaughnessy:2012oew}
O'Shaughnessy, R. 2013, Phys. Rev., D88, 084061

\bibitem[{{O'Shaughnessy} {et~al.}(2005){O'Shaughnessy}, {Kim}, {Fragos},
  {Kalogera}, \& {Belczynski}}]{Richard2005ApJ}
{O'Shaughnessy}, R., {Kim}, C., {Fragos}, T., {Kalogera}, V., \& {Belczynski},
  K. 2005, \apj, 633, 1076

\bibitem[{O'Shaughnessy {et~al.}(2008)O'Shaughnessy, Kim, Kalogera, \&
  Belczynski}]{Richard_2008}
O'Shaughnessy, R., Kim, C., Kalogera, V., \& Belczynski, K. 2008, The
  Astrophysical Journal, 672, 479

\bibitem[{{Petit} {et~al.}(2017){Petit}, {Keszthelyi}, {MacInnis}, {Cohen},
  {Townsend}, {Wade}, {Thomas}, {Owocki}, {Puls}, \& {ud-Doula}}]{Petit2017}
{Petit}, V., {Keszthelyi}, Z., {MacInnis}, R., {et~al.} 2017, \mnras, 466, 1052

\bibitem[{{Sadowski} {et~al.}(2008){Sadowski}, {Belczynski}, {Bulik},
  {Ivanova}, {Rasio}, \& {O'Shaughnessy}}]{Sadowski2008}
{Sadowski}, A., {Belczynski}, K., {Bulik}, T., {et~al.} 2008, \apj, 676, 1162

\bibitem[{Stevenson {et~al.}(2015)Stevenson, Ohme, \&
  Fairhurst}]{Stevenson:2015bqa}
Stevenson, S., Ohme, F., \& Fairhurst, S. 2015, Astrophys. J., 810, 58

\bibitem[{Stevenson {et~al.}(2017)Stevenson, Vigna-Gómez, Mandel, Barrett,
  Neijssel, Perkins, \& de~Mink}]{Stevenson:2017tfq}
Stevenson, S., Vigna-Gómez, A., Mandel, I., {et~al.} 2017, arXiv:1704.01352,
  [Nature Commun.8,14906(2017)]

\bibitem[{{Tutukov} \& {Yungelson}(1993)}]{Tutukov1993}
{Tutukov}, A.~V., \& {Yungelson}, L.~R. 1993, \mnras, 260, 675

\bibitem[{Voss \& Tauris(2003)}]{Voss:2003ep}
Voss, R., \& Tauris, T.~M. 2003, Mon. Not. Roy. Astron. Soc., 342, 1169

\bibitem[{{Xu} \& {Li}(2010)}]{Xu_Li_2010}
{Xu}, X.-J., \& {Li}, X.-D. 2010, \apj, 716, 114

\bibitem[{{Zevin} {et~al.}(2017){Zevin}, {Pankow}, {Rodriguez}, {Sampson},
  {Chase}, {Kalogera}, \& {Rasio}}]{Zevin2017}
{Zevin}, M., {Pankow}, C., {Rodriguez}, C.~L., {et~al.} 2017, ArXiv e-prints,
  arXiv:1704.07379

\end{thebibliography}
\end{document}